\documentclass[prl,aps,twocolumn,amsfonts,superscriptaddress,showpacs,floatfix]{revtex4}

\usepackage{color}
\usepackage{graphicx} 
\usepackage[tight]{subfigure}
\usepackage{multirow}
\usepackage{amsmath}

\begin{document}

\title{Constructing the generalized Gibbs ensemble after a quantum quench}

\author{Jean-S\'ebastien Caux}
\affiliation{Institute for Theoretical Physics, University of Amsterdam, Science Park 904, Postbus 94485, 1090 GL
Amsterdam, The Netherlands}

\author{Robert M. Konik}
\affiliation{CMPMS Dept., Brookhaven National Laboratory, Upton, NY 11973-5000, USA}

\date{\today}

\begin{abstract}
Using a numerical renormalization group based on exploiting an underlying exactly solvable nonrelativistic theory,
we study the out-of-equilibrium dynamics of a 1D Bose gas (as described by the Lieb-Liniger model) 
released from a parabolic trap.  Our method allows us to track the post-quench dynamics of the gas 
all the way to infinite time.  
We also
exhibit a general construction, 
applicable to all integrable models, of the thermodynamic ensemble that has been suggested to govern this dynamics, the 
generalized Gibbs ensemble.
We compare the predictions of equilibration from this ensemble against the long time dynamics observed using our method.
\end{abstract}


\maketitle

Understanding non-equilibrium quantum quench behavior in low-dimensional systems is a difficult theoretical challenge.
Because one is initializing the system in a state that is not an eigenstate, 
this behavior is determined not merely by the system's ground state (or a small number of excited states),
but rather by some coherent sum of a large number of eigenstates.  If one wants to explore the emergence of
a resulting steady state, the time evolution of this coherent sum must then be tracked over long periods of time.
This problem confronts theorists who wish to understand dynamics in perturbed quantum gases \cite{weiss,hofferberth}, 
ultrafast phenomena in 
superconductors \cite{bozovic}, and questions of thermalization in integrable systems \cite{rigol}.

This last set of questions arise because of the surprising experimental finding that a 
perturbed one-dimensional Bose gas retains
memory of its initial non-equilibrium state over long periods of time \cite{weiss} and does 
not appear to relax to a state of thermodynamic
equilibrium.  To understand this, it was proposed \cite{rigol}
that equilibriation does occur but not as described by a grand canonical ensemble (GCE).  Instead
the ensemble describing equilibriation needs to take into account the additional, non-trivial conserved 
quantities that, at least according
to the theoretical minimal model of the gas (the Lieb-Liniger (LL) model \cite{ll}), are present in the system.  
This new ensemble has been dubbed the
generalized Gibbs ensemble (GGE).  The GGE takes as the density matrix 
\begin{equation}
\hat\rho_{GGE} = Z^{-1}\exp (-\sum_i\beta_i Q_i)
\label{eq:rhoGGE}
\end{equation} 
where the $Q_i$ form an independent, complete
sequence of conserved quantities in the system and $\beta_i$ correspond
to a set of generalized (inverse) temperatures.  Computation of this
density matrix is non-trivial and has only been successfully accomplished in certain special limits.  
Most of these limits are in models 
where interactions (though not necessarily
correlation functions) correspond to a free model
(the hard core limit of the interacting
Bose gas \cite{rigol}, quadratic Hamiltonians \cite{barthel}, Luttinger liquids \cite{caza}, the sine-Gordon model
at the free-fermion point and in the semi-classical limit \cite{caza1}, and the quantum Ising model in the absence of a 
longitudinal field \cite{essler,rossini}).  A notable exception was 
the study of Fioretto and Mussardo \cite{fior_muss} where
it was possible to study quenches in general interacting integrable models but with the restriction to a very special
set of quench protocols.

It is against this backdrop that we present a general methodology able to study non-equilibrium behavior and quench dynamics
of low-dimensional interacting models, both integrable and non-integrable.  This method is predicated on a numerical 
renormalization group (NRG)
able to study models which can be represented as perturbed integrable and conformal field theories (CFT) \cite{ka1}:
\begin{equation}
H = H_{Integrable/CFT} + V_{perturbation}.
\end{equation}
The LL model
in a trapping potential takes this form.  We believe that this methodology is a valuable addition to other general methodologies
used to study dynamics in low-dimensional systems such as the time-dependent density matrix renormalization 
group \cite{2004_Vidal_PRL_93,2004_While_PRL_93,2004_Daley_JSTAT_P04005,kollath,schollwock}.
At least for a subset of quenches, where we quench into an integrable system (say by turning off the trapping potential in a LL
system), we can track the dynamics for all times.

Concomitant with the introduction of this tool to study quench dynamics, 
we present a general methodology to compute the density matrix
of the GGE using information arising from the application of the NRG.  
We show how one can write down a simple set of equations governing
the GGE and how the entire infinite set of generalized temperatures, $\{\beta_i\}^\infty_{i=1}$ 
can be readily determined.

The specific example we consider is the LL model perturbed by a one-body parabolic trap $V(x)=m\omega^2x^2/2$, 
\begin{equation}
H = -\frac{\hbar^2}{2m}\sum^N_{j=1}\frac{\partial^2}{\partial x_j^2}+2c\sum_{\langle i,j\rangle}\delta(x_i-x_j) + \sum_i V(x_i),
\end{equation}
(we will work in units where $2m=\hbar=1$).  In running the NRG, we use the basis of eigenstates
of the LL model and their matrix elements with respect to the trapping potential.  
Both the description of the states and the computation of matrix elements in the LL model
are much more complicated than the examples of relativistic field theories where the NRG has been applied previously.
The states in the LL model consist of N strongly interacting particles and 
not few-particle excitations above the true vacuum state, while the matrix elements do not see a chiral
factorization as in a relativistic gapless theory but are N-dimensional determinants \cite{1989_Slavnov_TMP_79_82}.  
To tackle this, we took recourse to a highly optimized set of routines known as ABACUS \cite{caux}
which solves and evaluates all equations needed to characterize both the necessary 
eigenstates and their matrix elements.   This package 
has been shown to be able to successfully compute dynamical response functions for
the LL model \cite{caux}.

We first use the NRG to extract the ground state of the LL model in a trap \cite{supp_mat}.
The NRG produces the ground state of the gas, $|\psi\rangle_{GS}$, as a linear combination of exact eigenstates, $|s\rangle$, 
of the LL model:
$
|\psi\rangle_{GS} = \sum_s c_s |s\rangle.
$
In order to accurately describe the ground state in the NRG procedure we typically
consider on the order of $10^4-10^5$ states.
We then consider a sudden release of the trap,
that is we will study the gas where we quench into
an integrable model.  For these types of quenches 
our methodology gives us the ability to study the evolution of the gas for arbitrary times. 
Each state, $|s\rangle$, appearing in the ground state
is characterized by a set of N (one for each particle) rapidities (quasi-momenta) $\{\lambda_n\}^N_{n=1}$.
These rapidities are solutions to the Bethe equations,
\begin{equation}\label{Beqn}
e^{i\lambda_nL} = \prod_{m\neq n} \frac{\lambda_n-\lambda_m+ic}{\lambda_n-\lambda_m-ic},
\end{equation}
and can be readily obtained to arbitrary accuracy.  
With the NRG we can compute the coefficients $c_s$  
with reasonably high accuracy \cite{supp_mat}.  Time evolution under the post-quench 
Hamiltonian (the unperturbed LL model) is extremely simple.  If $E_s$ is the energy of state $|s\rangle$, the time evolution is described
by 
$
|\psi(t)\rangle_{GS} = \sum_s c_s e^{-iE_st}|s\rangle.
$
Because each state's energy, $E_s$, is given in terms of the $\lambda_n$'s as $\sum_n\lambda_n^2$, we can compute the 
phases appearing in the above sum to arbitrary
accuracy for arbitrary time.  

To characterize the evolution of the gas in the long time limit 
we compute the 
momentum distribution function (MDF)
$n_k = \langle \psi^\dagger_k\psi_k\rangle$
in the diagonal ensemble (DE).
An observable ${\cal O^\dagger O}$ 
in this ensemble is simply given by 
\begin{equation}
\langle {\cal O^\dagger O}\rangle_{DE} \equiv \sum_s |c_s|^2 \langle s|{\cal O^\dagger O}|s\rangle.
\label{eq:corrDE}
\end{equation}
To compute this correlation function we insert a resolution of the identity between ${\cal O^\dagger}$ and ${\cal O}$ and use
a specially-designed version of ABACUS for excited states to compute all of the necessary matrix elements \cite{supp_mat}.

\begin{figure}
\includegraphics[width=0.4\textwidth]{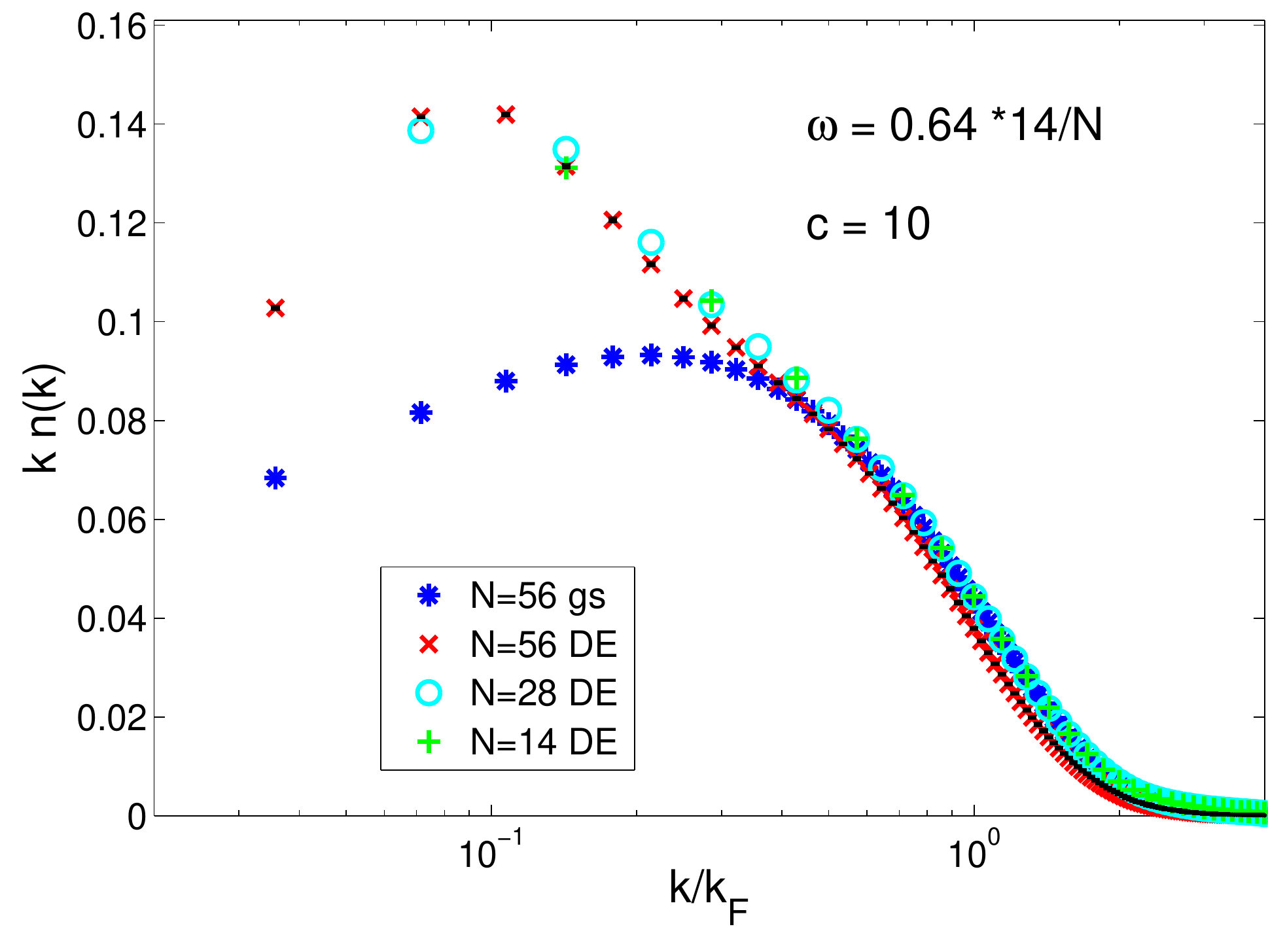} \\
\includegraphics[width=0.4\textwidth]{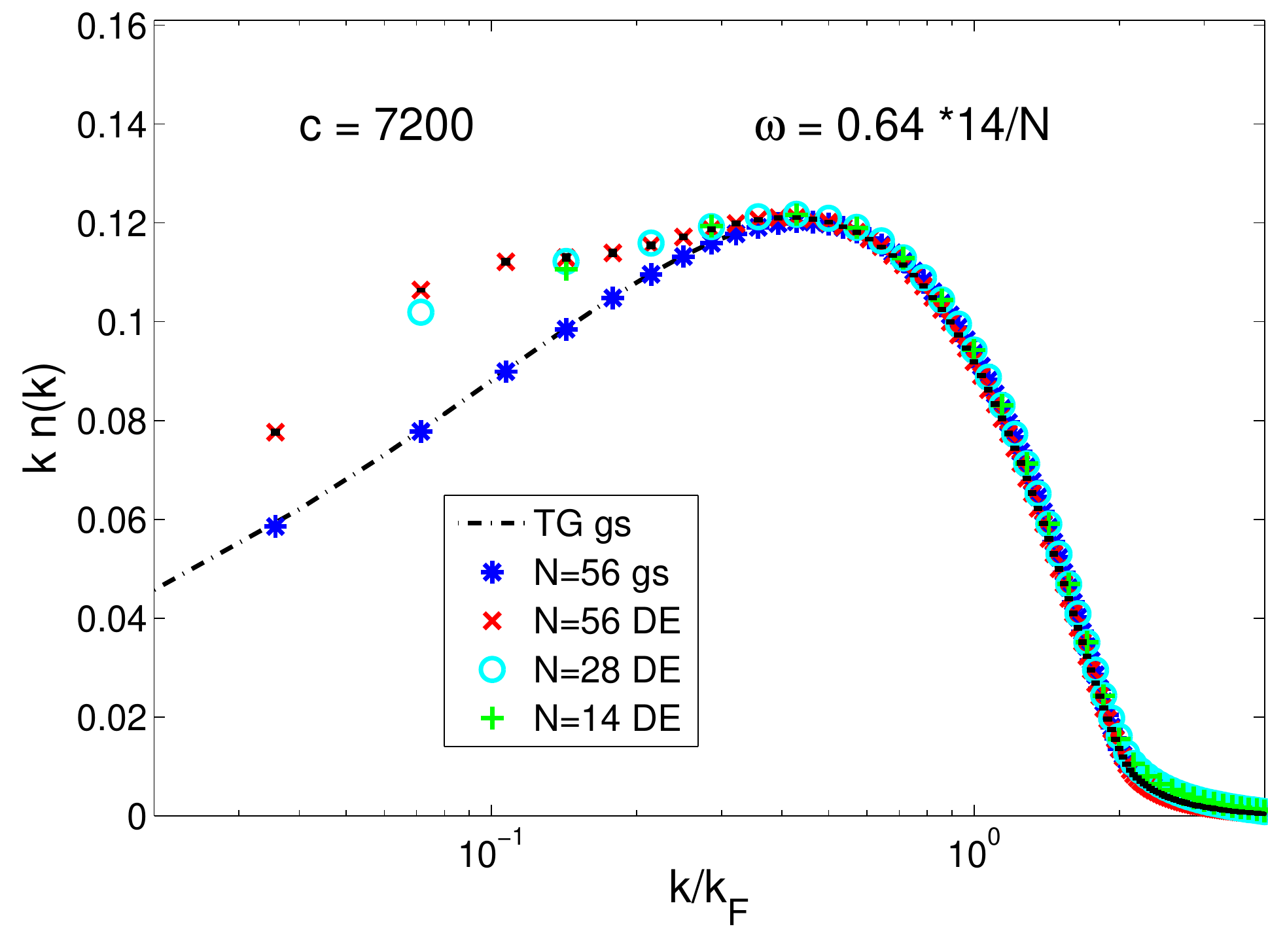}
\caption{The MDF in the DE of the gas after release from a trap 
for $c = 10$ (top) and $c=7200$ (bottom).  Shown are
the gases at ($N=L=14, \omega=0.64$), ($N=L=28,\omega=0.32$), and ($N=L=56,\omega=0.16$).
Error bars are given for the $N=L=56$ data alone and are estimated from the speed of convergence of the NRG (see \cite{supp_mat}) --
we believe the $N=L=14,28$ data is completely converged.
The MDF of the untrapped gas ($N=L=56$) is shown for comparison as is the analytic expression
available for the Tonks-Girardeau gas $c=\infty$ from Ref. \cite{lenard}.}
\label{fig:MDF}
\end{figure}

In Fig. \ref{fig:MDF} we plot the MDF in the DE of the gas post-release for two values of c ($c=10$ and $c=7200$) and 
for a variety of system sizes, with $\omega L$ fixed and keeping $N=L$.
For comparison we also plot the MDF of the gas in its ground state. 

We see, as expected, that the MDF of the gas is perturbed from
that of the ground state at low momenta but remains unchanged from the ground state MDF 
at higher momenta.  The relative insensitivity to different values of $N=L,\omega$ is consistent
with a perturbative (in $\omega$) computation of the MDF in the DE at $c=\infty$ which shows 
$n(k)_{DE}= n(k)_{GS}+(\frac{\omega L}{2\pi})^4\frac{N}{L}\frac{m^2\sqrt{2\pi}B_0}{8v_F^2k^{5/2}} +{\cal O}(\omega^8)$.  Here
$n(k)_{GS}$ is the MDF of the ground state, the constant $B_0\approx0.5124$ \cite{adilet}, and $v_F$ is the velocity
of the gas.  The scaling with $N,L$, and $\omega$ indicated by this expression implies that
variations in $n(k)_{DE}$ between different system sizes in Fig. 1 are due to finite size corrections 
which are small (on the order of the symbol size).
As an important check of our results, the high momenta tails of the MDF's at $c=7200$ 
behave as the predicted $k^{-4}$ \cite{lenard,2003_Olshanii_PRL_91,2003_Gangardt_PRL_90}.

While the diagonal ensemble tells us what the final steady state of the gas is after its release, a question of primary interest
is whether the steady state can be associated with some ensemble.  It has been postulated \cite{rigol}
that for a quench into an integrable system the correct ensemble to use is the GGE ensemble in Eqn. \ref{eq:rhoGGE}.  
The $Q_i$'s are here non-trivial polynomials in the field operators (and their derivatives) \cite{korepin}.  
The action of the $Q_i$'s on the states, $|s\rangle$ is straightforward.  With each
state, $|s\rangle$, characterized by a set of $N$ rapidities, $\lambda_i$, the action of the $Q_i$ upon $|s\rangle$ is
$
Q_i|s;\lambda_1,\cdots,\lambda_N\rangle = \sum_j \lambda_j^i|s;\lambda_1,\cdots,\lambda_N\rangle,
$
that is to say, $Q_i$ acts on the state like an i-th power sum.
This shows that the $Q_i$'s are both a complete and independent
set of charges inasmuch as the polynomials form a complete and independent basis in the space of single variable functions.

To compute $\hat\rho_{GGE}$ the most straightforward path is to compute $\langle Q_i\rangle$ at $t=0$ 
and insist that the set of $\beta_i$'s is such that ${\rm Tr}(\hat\rho_{GGE}Q_i)$ gives the
same answer.  In the case of the hard core limit this is readily doable as the $Q_i$'s can be written in terms of
a more amenable basis, the momentum occupation numbers:
$
Q_i = \sum_\lambda \lambda^i n_\lambda,
$
where $n_\lambda$ tells you whether there is a particle with rapidity of the form $\lambda = 2\pi m/L$ for $m\in \mathbb{Z}$.
In this basis of charges, $\langle n_\lambda \rangle_{GGE}$ simplifies to 
${\rm Tr}(\exp(-\beta_\lambda n_\lambda)n_\lambda)/{\rm Tr} \exp(-\beta_\lambda n_\lambda)$, i.e. for such expectation values the ensemble
factorizes, and $\beta_\lambda$ is readily computed.  This simplification, however, does not exist away from the hard core
limit and we are instead left with a complicated non-linear minimization problem which on the face of it does not obviously
have a solution.  We now show that it does and that the $\beta_i$'s can be computed readily.  We do so through a (generalized)
thermodynamic Bethe ansatz \cite{jorn}.

Because the action of the charges $Q_i$ on the states, $|s\rangle$,
are given simply in terms of the rapidities, $\lambda_i$, identifying 
the state, to ask that $\langle Q_i\rangle_{t=0} = \langle Q_i\rangle_{GGE}$ amounts to asking whether there is a set of $\lambda$'s,
$\{\tilde\lambda_j\}^N_{j=1}$, such that
$$
\langle Q_i \rangle_{t=0} = \sum_j \tilde\lambda^i_j, ~~~ i=1,2,\cdots.
$$
There is in fact such a set.  We can moreover determine its rapidity distribution, which we will call $\rho_{GGE}(\lambda)$, 
directly from
$|\psi\rangle_{GS}$.  To each state, $|s;\lambda_{s1},\cdots,\lambda_{sN}\rangle$, we associate a distribution, $\rho_s(\lambda)$, governing
the $\lambda$'s of that particular state:
$
\rho_s(\lambda) = \frac{1}{L}\sum_i \delta(\lambda -\lambda_{si}).
$
Then $\rho_{GGE}(\lambda)$ is the weighted sum of the $\rho_s(\lambda )$'s:
$$
\rho_{GGE}(\lambda) = \sum_s |c_s|^2 \rho_{s}(\lambda ).
$$
In particular $\int d\lambda \rho_{GGE}(\lambda )\lambda^i = L\langle Q_i\rangle_{t=0}$.

\begin{figure}
\includegraphics[width=0.35\textwidth,height=0.28\textwidth]{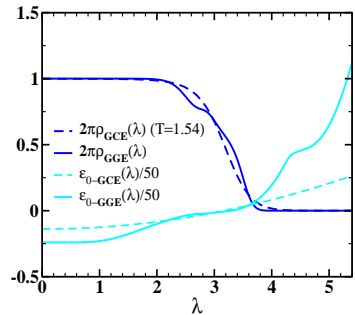}
\caption{$\varepsilon_{0}(\lambda )$ and $\rho(\lambda )$ for both the GGE and GCE ensembles for a gas with $N=L=56$, $c=7200$,
and a prequench trap strength, $\omega=0.256$.  For the GCE ensemble the effective temperature is $T=1.54$.  The quantities
plotted are symmetric about $\lambda=0$.}
\label{fig:rho}
\end{figure}

$\rho_{GGE}$ contains,
implicitly, all the information to characterize the action of $\hat\rho_{GGE}$ on a eigenstate of the LL model \cite{jorn}.
A distribution of $\lambda$'s must be consistent with the Bethe equations (Eqn. \ref{Beqn}).  In the continuum limit, these
equations can be rewritten as \cite{ll,yy}
\begin{equation}\label{den_eqn}
\rho_{GGE}(\lambda) + \rho^h_{GGE}(\lambda) = \frac{1}{2\pi} + \int \frac{d\lambda'}{2\pi} K(\lambda-\lambda')\rho_{GGE}(\lambda),
\end{equation}
where $\rho^h_{GGE}(\lambda)$ is the density of holes in the $\lambda$-distribution and $K(\lambda)=2c/(c^2+\lambda^2)$.  Now the GGE is
derived by the same principles as the grand canonical ensemble: namely entropy is maximized subject to the constraints of fixed
conserved charges (energy for the grand canonical ensemble, all the charges, $Q_i$, for the GGE).  Thus associated with GGE 
is a generalized free energy
$
F_{GGE} = \int d\lambda\rho_{GGE}(\lambda)\varepsilon_{0-GGE}(\lambda) - S,
$
where $\varepsilon_{0-GGE}(\lambda) \equiv \sum_i \beta_i \lambda^i$ is 
a generalized energy.   It corresponds to the action of
$\hat\rho_{GGE}$ on a state $|s;\lambda_1,\cdots,\lambda_N\rangle$:
\begin{equation}
\hat\rho_{GGE}|s;\lambda_1,\cdots,\lambda_N\rangle = \frac{e^{-\sum_i\varepsilon_{0-GGE}(\lambda_i)}}{Z}|s;\lambda_1,\cdots,\lambda_N\rangle.
\end{equation}
In particular 
knowing $\varepsilon_{0-GGE}$ then allows us to compute general expectation values in the GGE.
While $\varepsilon_{0-GGE}$ differs from its form in the grand canonical ensemble, 
$S$ is the standard entropy \cite{yy} of a system 
with a given distribution of particles, $\rho_{GGE}$, and holes, $\rho^h_{GGE}$:
\begin{eqnarray}
S &=& \int d\lambda \bigg[ (\rho_{GGE}+\rho^h_{GGE})\log(\rho_{GGE}+\rho^h_{GGE})\cr
&& \hskip -.25in - \rho_{GGE}\log\rho_{GGE}-\rho^h_{GGE}\log\rho^h_{GGE}\bigg].
\end{eqnarray}
We now show that we can express $\varepsilon_{0-GGE}$ in terms of $\rho_{GGE}$ that we derived from $|\psi\rangle_{GS}$.

If we minimize the generalized free energy we arrive at a constraint between the particle and hole distributions and $\varepsilon_{0-GGE}$:
\begin{eqnarray}\label{en_eqn}
\varepsilon(\lambda ) \!=\! \varepsilon_{0-GGE}(\lambda) \!-\!\int \frac{d\lambda'}{2\pi}K(\lambda\!-\!\lambda')\log(1+e^{-\varepsilon(\lambda)}),
\end{eqnarray}
where $\varepsilon = \log(\rho^h_{GGE}/\rho_{GGE})$.  Thus to determine $\varepsilon_{0-GGE}$ we take our knowledge of $\rho_{GGE}(\lambda)$
obtained from $|\psi\rangle_{GS}$, use Eqn. (\ref{den_eqn}) to determine $\rho^h_{GGE}$ which then gives us $\varepsilon(\lambda)$.  From
Eqn. (\ref{en_eqn}), we then can fix $\varepsilon_{0-GGE}$.  

\begin{figure}
\vskip 10pt
\includegraphics[width=0.4\textwidth]{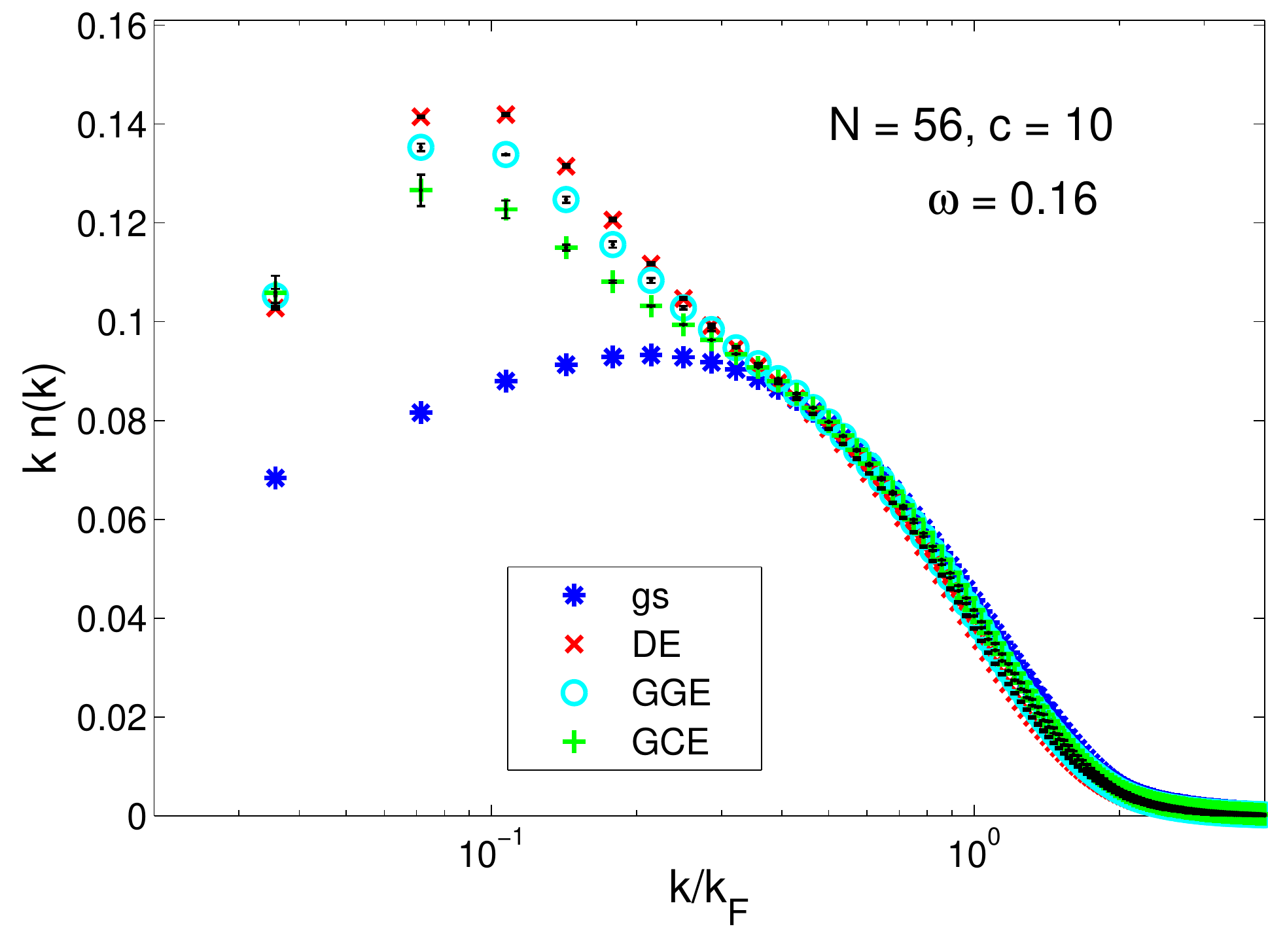} \\
\includegraphics[width=0.4\textwidth]{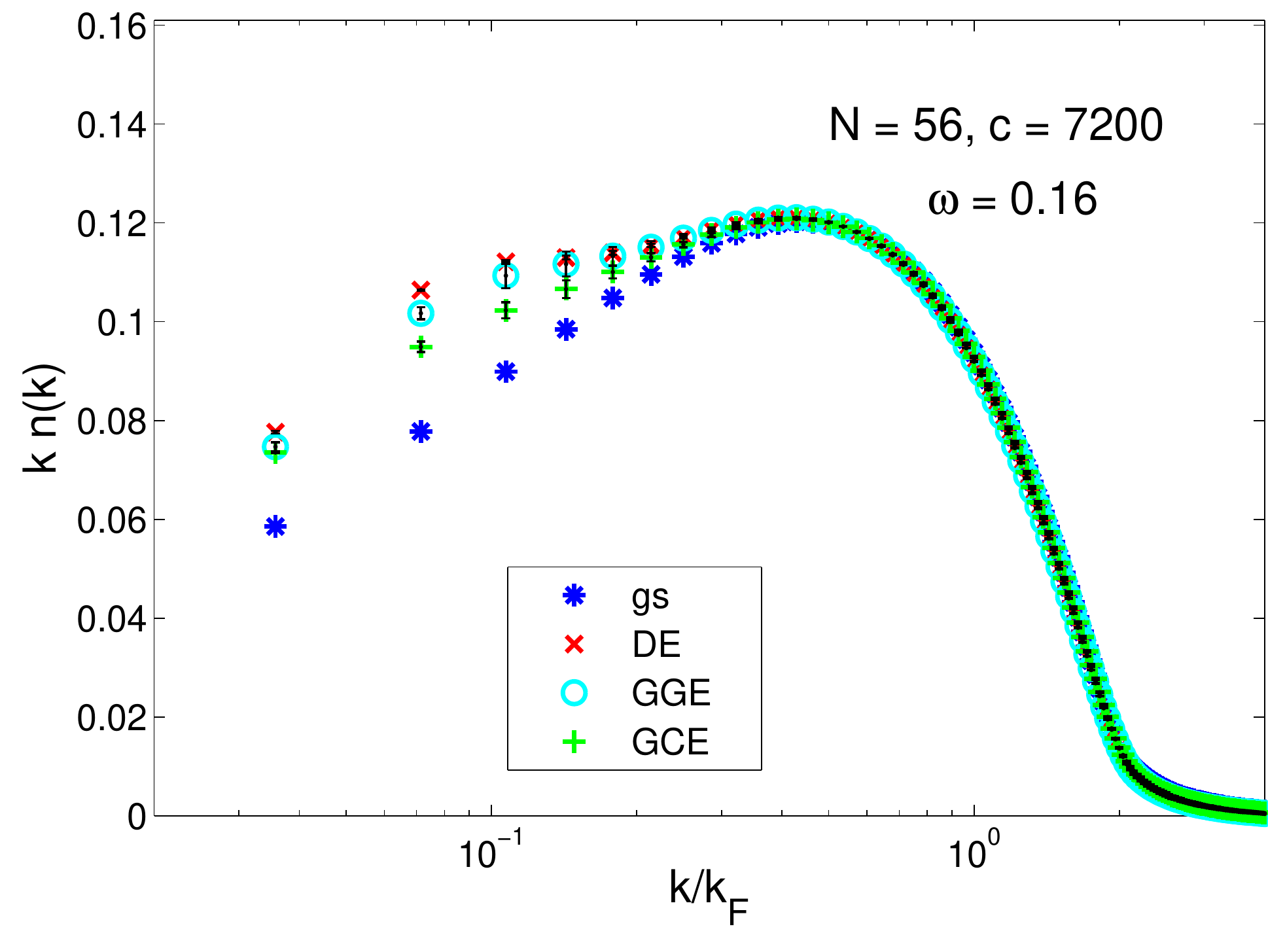}
\caption{The MDF (for $N=L=56$) in the GCE and GGE for the gas after release from a trap of strength $\omega=0.16$
for $c = 10$ (top) and $c=7200$ (bottom).  
We again show the MDF of the untrapped gas for comparison (blue stars).}
\label{fig:MDFGE}
\end{figure}

Following this procedure we plot in Fig. \ref{fig:rho} $\rho_{GGE}$ and $\varepsilon_{0-GGE}$ for the gas in the hard core limit.  
For comparison we plot
what these quantities would be if instead of a generalized Gibbs ensemble, the thermodynamics was governed by the grand canonical
ensemble.  (In this case we use the standard thermodynamic Bethe ansatz equations \cite{yy} to determine what $\rho_{GCE}$ and 
$\varepsilon_{0-GCE}=\beta(\lambda^2-\mu)$ need to be, i.e. what the effective temperature needs to be,
if they are to reproduce the correct density and average energy of the system $_{GS}\langle \psi|H|\psi\rangle_{GS}$.)
We see that both $\rho_{GGE}$ and $\varepsilon_{0-GGE}$ have considerably more structure than that of their grand canonical
counterparts.

We now use this ability to compute $\varepsilon_{0-GGE}(\lambda)$, to compute various expectation values of observables 
in the GGE. In Fig. \ref{fig:MDFGE} 
we plot the MDF as computed in the DE and in both the GGE and GCE.  
The error estimate is computed similarly as in Fig. \ref{fig:MDF} (see \cite{supp_mat} for details).
For the data at hand, we see that for low momenta the two ensemble averages, GGE and GCE, disagree
with the DE.  However the GGE provides
a considerably better match to the DE than does the ordinary thermal ensemble GCE.  From the finite size comparison
(see Fig. 3 of \cite{supp_mat}, 
it can be argued (although not conclusively) that at small but finite $k$, this difference will vanish
with increasing system size.

The disagreement between ensembles in the data is not entirely surprising.  
The logic of the GGE is such that it is expected to describe correlations that are local in space (and that
involve a distance scale significantly smaller than the system size).  We thus do not expect the correlations
at $k\sim 1/L$ to be particularly well described by the GGE.  
However there is the possibility that the differences between ensembles will remain at finite $k>1/L$ even in the
infinite volume limit.
In recent work \cite{santos1} the entropy associated with the DE was shown to be considerably 
smaller than that of the GGE implying that the DE is more tightly constrained than the GGE, i.e. the GGE
seems to be missing correlations.  It would be interesting to understand if this missing entropy is solely associated
with non-local correlations.

In conclusion, we demonstrated how an NRG based on exploiting the integrability
of the LL model can be used to study the time-dependent evolution after a quantum quench where
a 1D gas is released from a parabolic trap.  We have also demonstrated how to use the information arising
from the NRG to construct the corresponding GGE which has been suggested as a possibility for governing the
post-quench dynamics.  While we have focused on the LL model, this methodology is applicable to
any non-relativistic integrable theory of which the Heisenberg and XXZ spin chains are two prominent
examples.

\vskip 10pt

\noindent Acknowledgements: This research was supported by the US DOE (DE-AC02-98CH10886),
the New York Center for Computational Sciences at Stony Brook University/Brookhaven National Laboratory,
the Foundation for Fundamental Research on Matter, and the
Netherlands Organisation for Scientific Research. We thank G. Brandino and J. Mossel for useful discussions.

\pagebreak
\appendix

\section{Supplementary Material}

\subsection{Description of the numerical renormalization group}

The NRG we employ is one appropriate to the study
of continuum field theories \cite{ka1,k2} which in turn is based upon Wilson's numerical
renormalization group first used to study quantum impurity problems \cite{wilson}.  The basic idea behind
Wilson's numerical approach is that one performs a series of numerical diagonalizations which 
are ordered such that ``important'' states in the considered Hilbert space of the problem are taken into account first,
while the effect of less important states are included only in subsequent diagonalizations.  The sequence of diagonalizations,
the sequence of renormalizations, are done such that the numerical burden is the same at each step in the sequence.
The metric that determines the order in which states are taken into account, however, is arbitrary and can be chosen to 
be appropriate for the problem
at hand.  In Wilson's case, namely the Kondo problem where the impurity spin sits at the end of a half line
lattice, states involving only the Kondo spin together with nearby lattice sites are taken into account first.
That such states are most significant is guaranteed by lattice hopping parameters that decrease in magnitude the
further one gets away from the impurity.  In the case of a quantum critical Ising model perturbed by a magnetic field
(considered in \cite{ka1}) the Hilbert space used to form the matrices in the sequence of numerical
diagonalizations is that of the quantum critical Ising model.  The states in this model are ordered in terms of energy
(relative to the unperturbed theory).  Here low-energy states are the most important as the spin operator coupling
to the magnetic field is highly relevant and so are taken into account first by the renormalization group.

However for models like the LL model in a trapping potential
the complexity of the eigenstates (as we have discussed in the main body of the text)
means that energy alone is not a sufficient metric for the NRG to distinguish important from 
less important states; using such a limited metric would make the procedure drastically sub-optimal.  
To overcome this problem we introduce a variational metric in the space of states similar to that used to compute the single
particle spectrum of semiconducting carbon nanotubes \cite{k2}.  
Although this latter problem could be represented as a perturbation 
of a conformal field theory (CFT),
the CFT was complex enough (four bosons) that the same issues arose.  This variational metric uses an iterative
process which amounts to performing successively higher order computations in perturbation theory to determine 
which states are 
likely to significantly contribute to the low-energy spectrum of the fully perturbed theory.

\begin{figure}
\subfigure{\includegraphics[width=0.26\textwidth,height=0.26\textwidth]{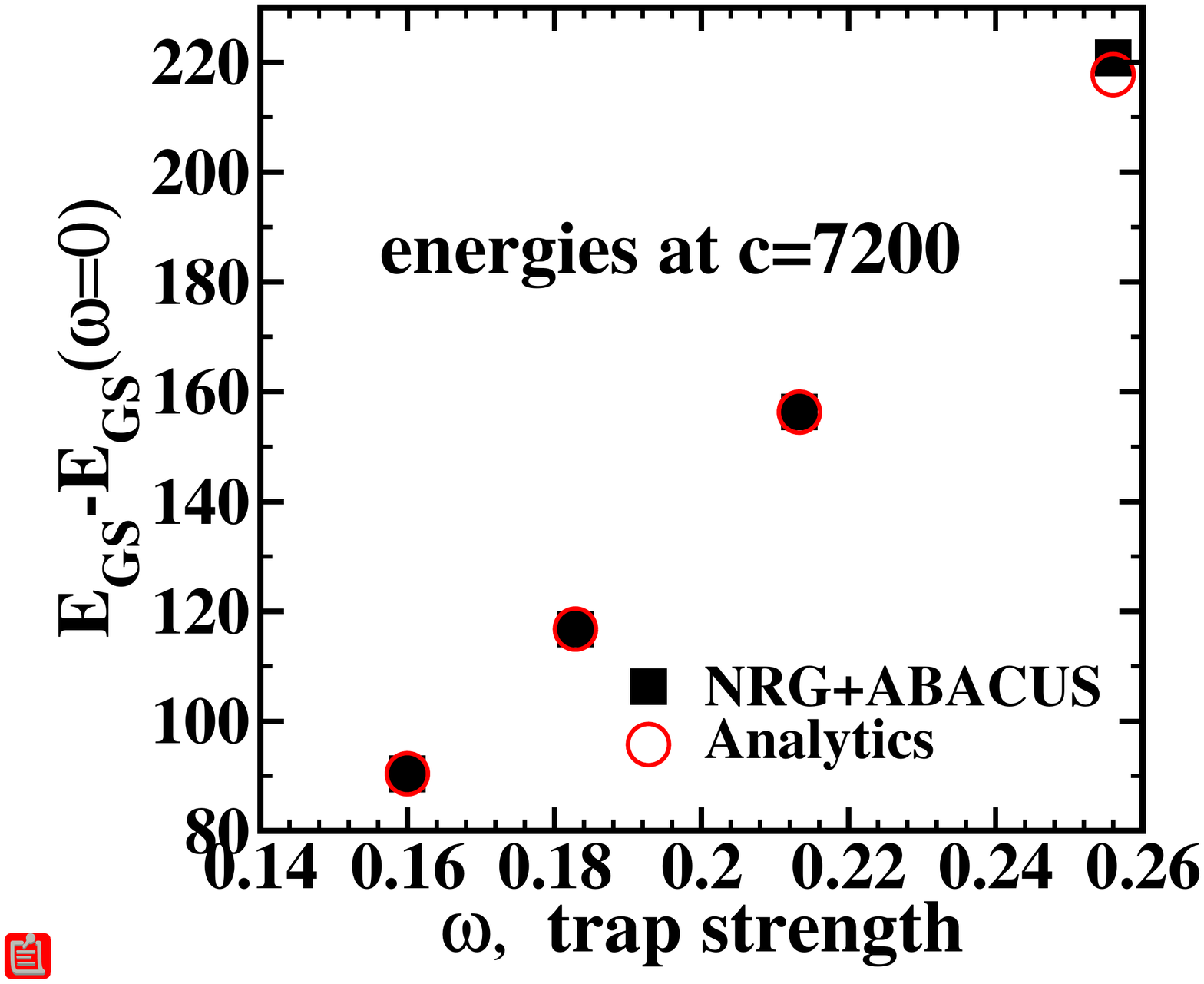}}\hskip -.28in
\subfigure{\includegraphics[width=0.26\textwidth,height=0.26\textwidth]{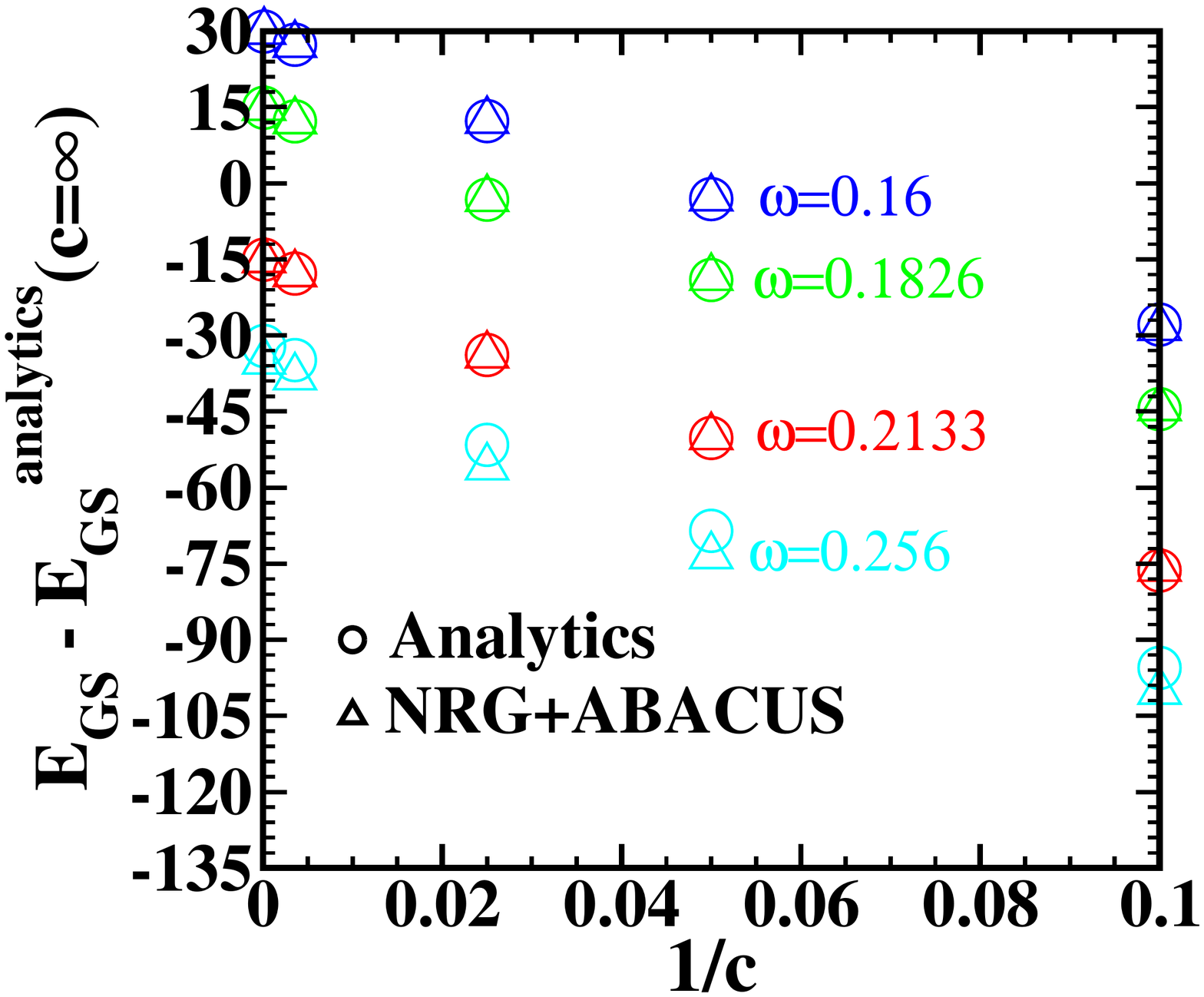}}
\caption{Comparing analytics with NRG+ABACUS for computation of the ground
state energy for different trap strengths and values of $c$ at $N=L=56$.  
Left: Ground state energy as a function of $\omega$. Right: Ground state
energy at different values of $c$ -- analytics include up to $1/c^3$ contributions.  Numerical data for different 
$\omega$ have been offset from each other.}
\label{fig:gs_energy}
\end{figure}

\subsection{Description of the gas in the trap in equilibrium}

While the primary concern of the accompanying letter was the discussion of correlations, 
we feel it is important here to provide
some details of the method so that readers can be reassured that we can describe the gas in the trap accurately,
that is to say, that we can provide a reasonable description of the pre-quench state.  We will provide
a more detailed description of these computations in a future publication \cite{unpublished}.

To demonstrate this we focus on the large $c$ limit where in the extreme Tonks-Girardeau limit ($c=\infty$) the system 
reduces to one of free fermions (at least for the computation of the energy) in a trap and where $1/c$ corrections 
can be systematically computed
by mapping the Bose system to one of fermions interacting with an ultra-short ranged potential with 
strength proportional to $1/c$ 
\cite{1999_Cheon_PRL_82}.

\begin{figure}[t]
\subfigure{\includegraphics[width=0.35\textwidth,height=0.25\textwidth]{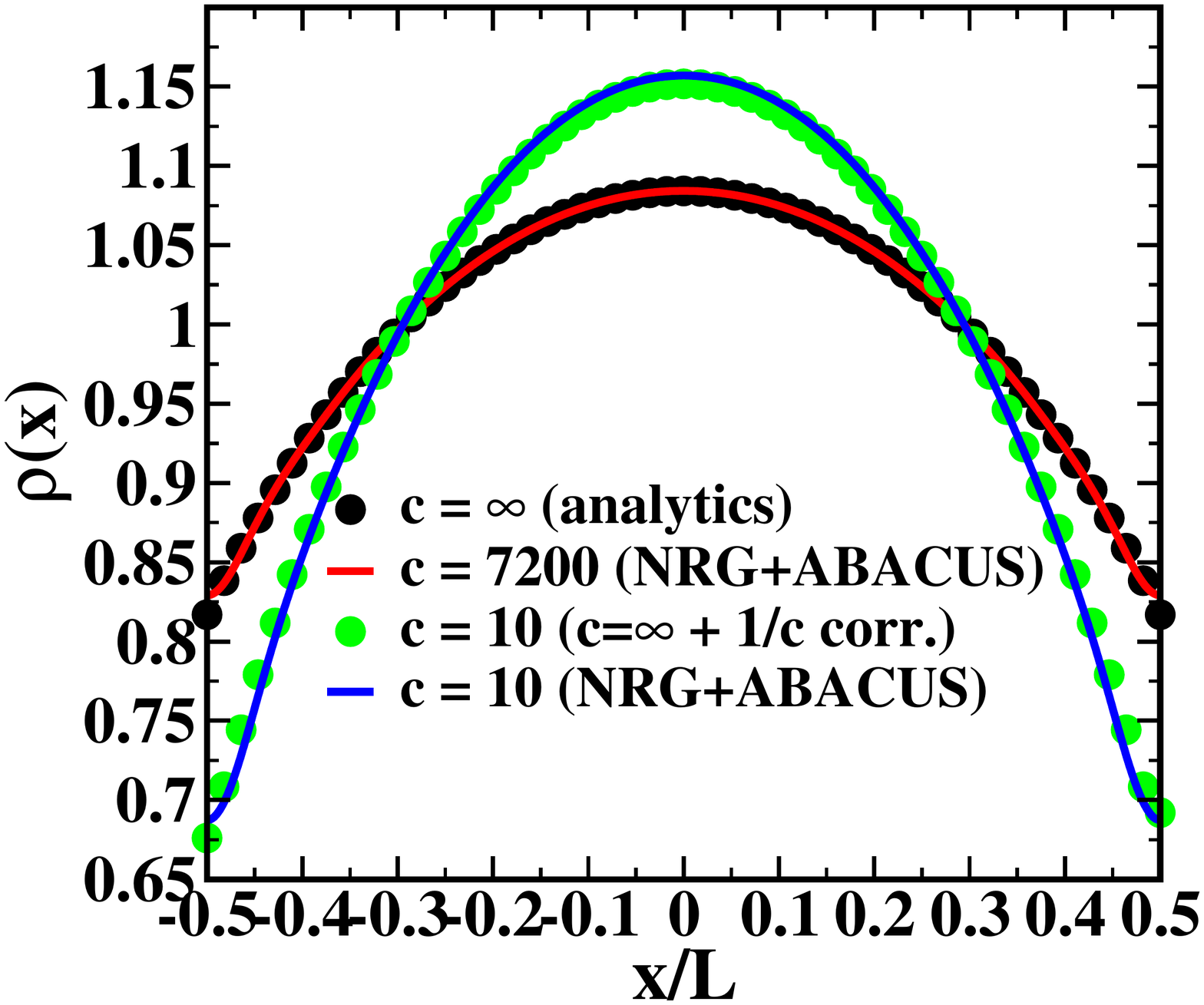}}\hskip -.2in
\subfigure{\includegraphics[width=0.35\textwidth,height=0.25\textwidth]{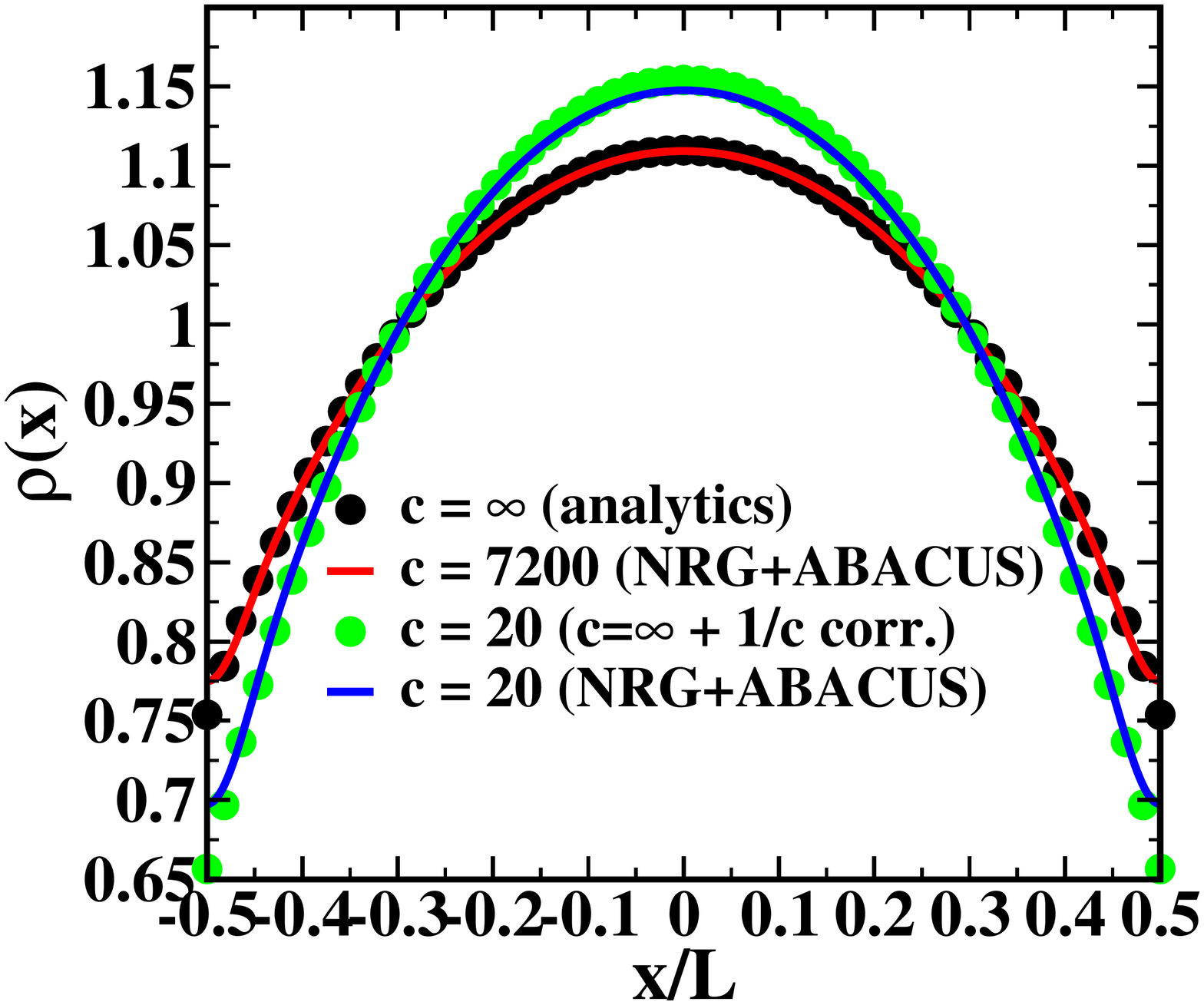}}
\caption{Density profiles in the trapped gas at two different values of $c$ comparing NRG+ABACUS 
with analytics for $N=L=56$: top) $\omega=0.16$; bottom) $\omega=0.183$.}
\label{fig:excited_states1}
\end{figure}

We present the computation of the ground state energies in Fig. 4.  In the left panel we present the ground state 
energies of the gas
(with $N=L=56$) in traps of different strengths, $\omega$.  
We get good agreement between the NRG computation and the analytics (better than 0.02\%
for the first three trap values and about 1\% for the largest of the trap values, $\omega=0.256$, studied).  In the 
right panel we show the
computation of the ground state energies as a function of $c$ for values running from $c=7200$ to $c=10$ for the 
same four values of the trap
strength.  In order to match analytics with numerics we needed to include corrections up to $1/c^3$.  
Computing the $1/c$ correction is straightforward.
The $1/c^2$ correction, when computed naively with second order perturbation theory, shows an ultraviolet divergence related to the
ultra short ranged potential
of the equivalent fermionic model.  This divergence can however be regulated with a point splitting procedure 
adopted from Ref. \cite{sen}.
However these first two corrections are not enough to get good agreement for the $c=10$ values of the ground 
state energies.  We thus
include the $1/c^3$ correction coming from the untrapped gas and which is readily computed from its integrability \cite{ll}.  
With these
three analytic contributions in place, we see we get excellent agreement for the first three trap values while the strongest trap
value ($\omega=0.256$) continues to see a deviation on the order of $1\%$ for all values of $c$.

Finally we consider our ability to accurately compute the density profile of the gas in the trap.  This is a much more 
complicated quantity to compute
than the ground state energies, since the matrix elements of the density operator must be employed.
In Fig. 5 we plot the density profile of the gas for two values of trapping strength and two values of $c$.
Again we work at larger values of $c$ where we can compare our numerics to an analytic computation ($c=\infty$ 
plus first order $1/c$ corrections).
We see that we get good agreement in all cases.

\begin{figure}[t]
\includegraphics[width=0.4\textwidth,height=0.35\textwidth]{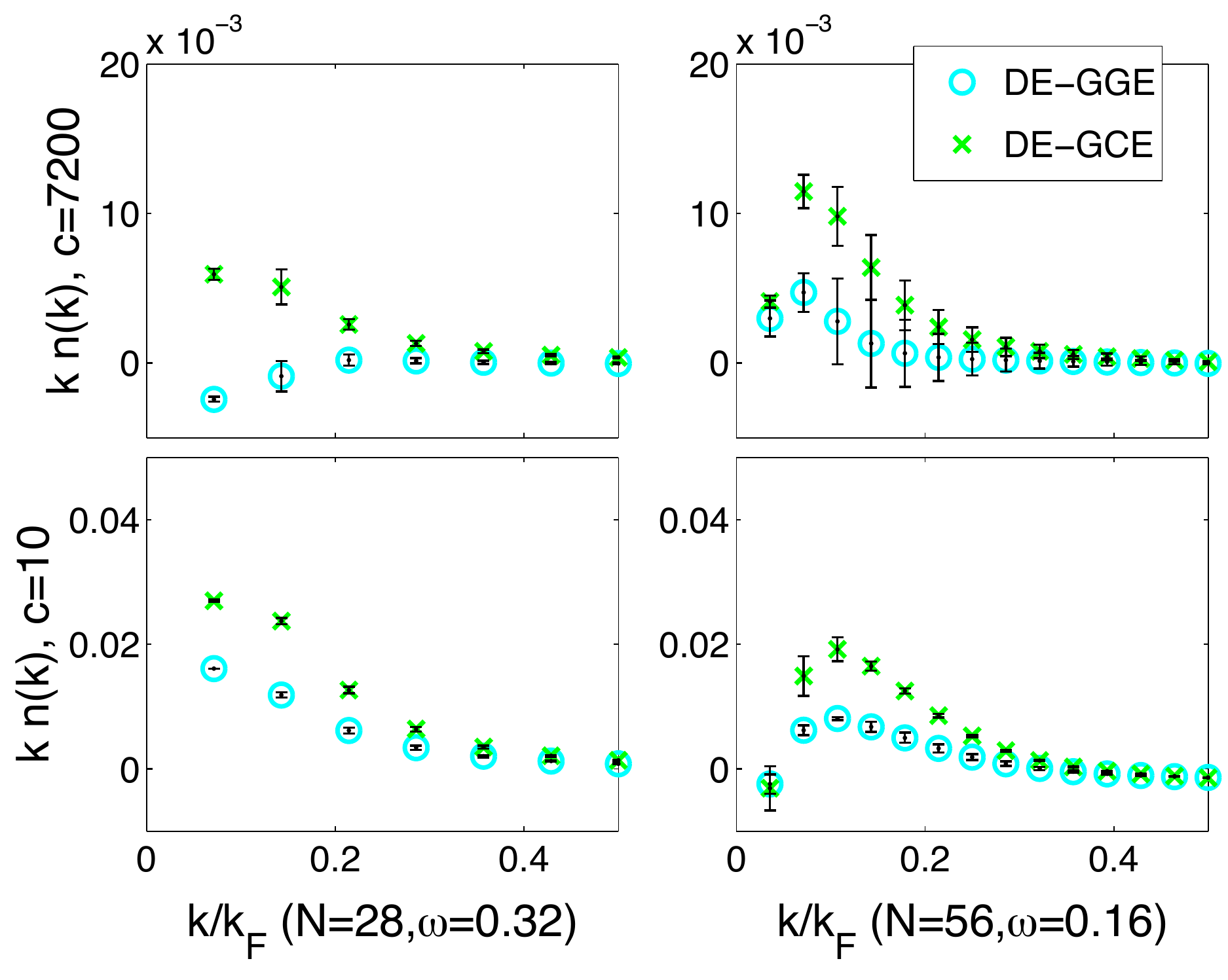}
\caption{The difference between the MDF as computed in the DE 
and as computed in the GGE and GCE for $N =L= 28$ (left) and $N = L=56$ (right).
Top: The gas after release from a trap of strength $\omega = 0.32~ (N=L=28)$ and $\omega = 0.16~ (N=L=56)$ for $c=7200$. 
Bottom: The gas after release from the same traps but for $c = 10$.
The GGE estimate is seen to be more accurate than the GCE throughout the range of values considered.}
\label{fig:MDFGE1}
\end{figure}

\subsection{Finite Size Corrections to the MDF}

In Fig. 6 we present data for system sizes $N=L=28$ and $N=L=56$ of the differences of the MDF between the GGE and the DE as well as
the GCE and the DE.
We see that generically differences are present for small momenta.  However these differences
seem to behave differently for the different ensembles.  For DE-GCE these differences remain approximately the same
as one moves from $N=L=28$ to $N=L=56$.  However for DE-GGE, these differences are notably reduced in increasing the system
size, particularly for the $c=10$ case.  While we do not have sufficient data to perform a finite size scaling analysis, it appears
that the difference DE-GGE is vanishing as system size grows, while the difference DE-GCE remains at some finite, though k-dependent, value.

\begin{figure}[t]
\includegraphics[width=0.4\textwidth]{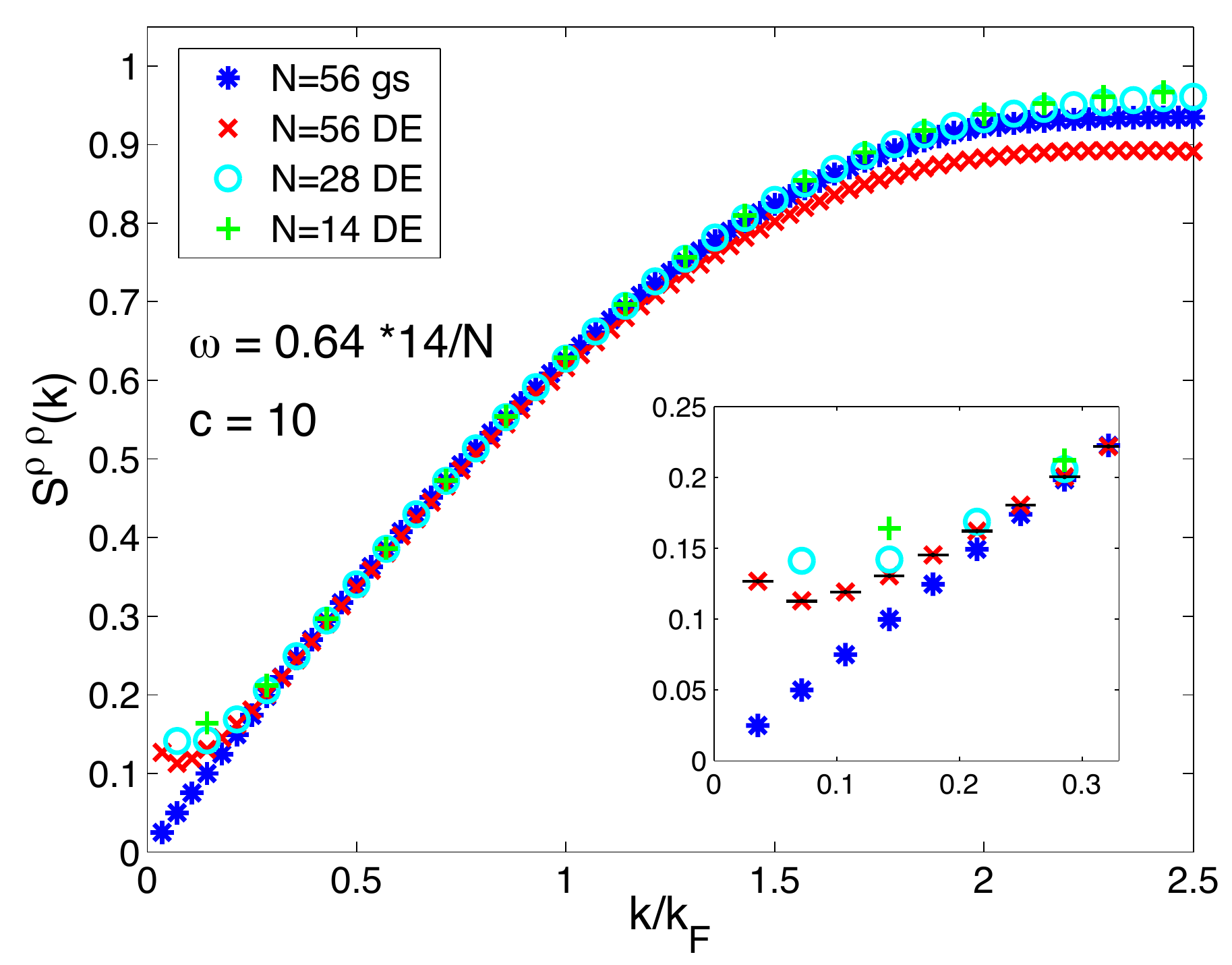} 
\includegraphics[width=0.4\textwidth]{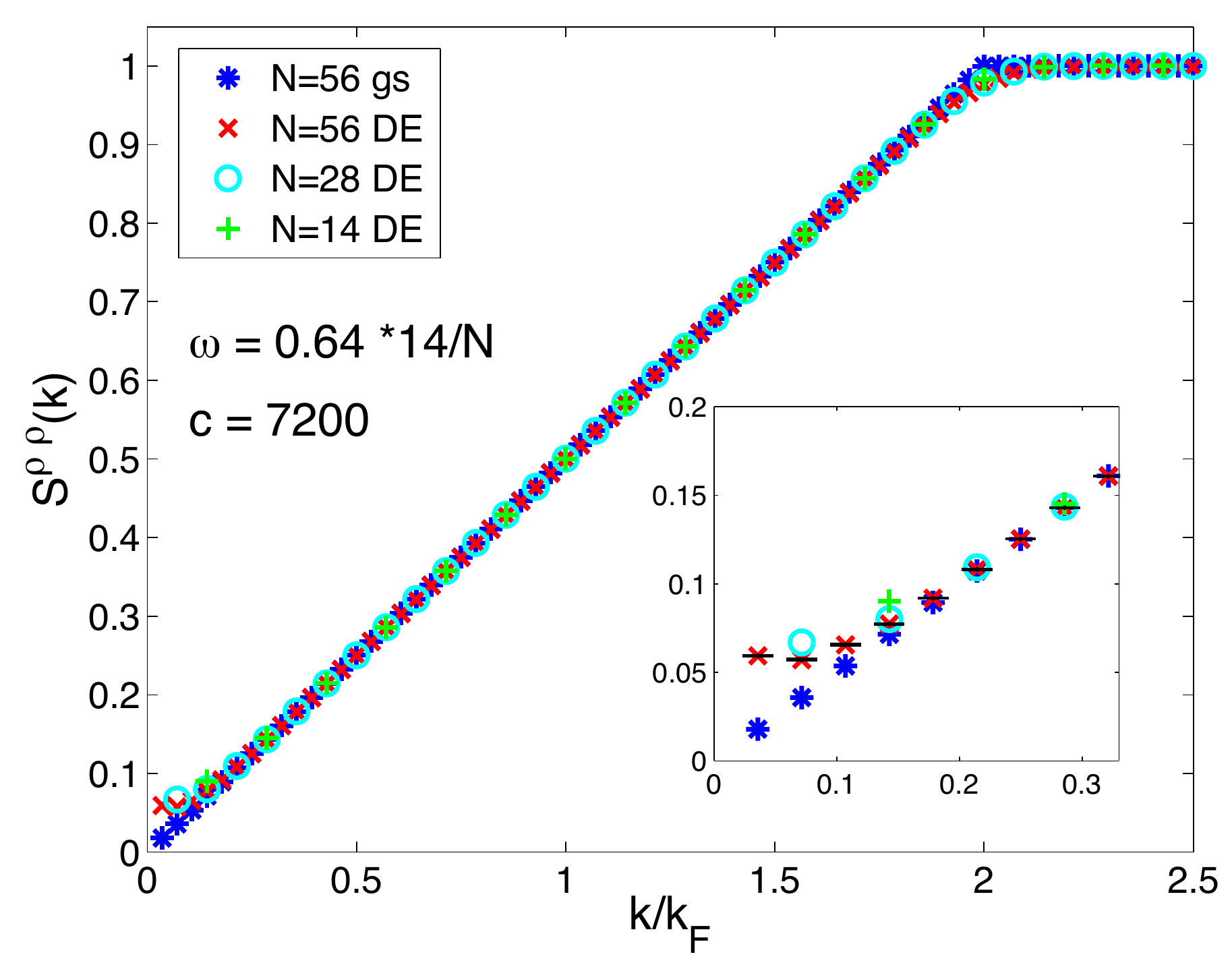} 
\caption{The SSF in the DE of the gas after release from a trap of strength $\omega$
for $c = 10$ (top) and $c=7200$ (bottom).  Shown are
the gases at ($N=L=14, \omega=0.64$), ($N=L=28,\omega=0.32$), and ($N=L=56,\omega=0.16$).
The $N=L=56$ data for higher momenta $k \sim 2k_F$ is not fully saturated.
The error bars (given in insets only) are calculated as for the MDF (see text).
}
\label{fig:SSF}
\end{figure}

\subsection{The SSF in the Various Ensembles}

While we focused on the MDF in the main body of the text,
we also have computed the 
density-density correlation function (static structure factor (SSF)) 
$S^{\rho \rho} (k) = \frac{1}{L}\langle \rho_k \rho_{-k}\rangle$
(with $\rho_k = \sum_q \psi^\dagger_{k+q}\psi_q$ -- for definitions of the density operator see Ref. \cite{cc}).
In Fig. \ref{fig:SSF} we plot the SSF in the DE for $c=10$ and $c=7200$.
We see that the effects of the trap are restricted to small momenta and that they are enhanced as one reduces $c$, moving
away from the hardcore limit.  

The role of scaling with system size is more complicated with the SSF than with the MDF.  With the MDF we were able to argue
(at least for weak values of the trap) 
that it remained invariant under the following scaling: $N,L,\omega \rightarrow 2N,2L,\omega/2$.  This is not true for the SSF.
At weak values of $\omega$, we find that the SSF in the DE is given by
\begin{equation}\label{srrde}
S^{\rho\rho}_{DE}(k\ll k_F) = S_{GS}(k) + \frac{m^4\omega^4}{\pi k_F^2 k^5} + {\cal O}(\omega^8).
\end{equation}
The simpler scaling for the MDF led us to emphasize this quantity in the main body of the text.

\begin{figure}
\includegraphics[width=0.4\textwidth,height=0.35\textwidth]{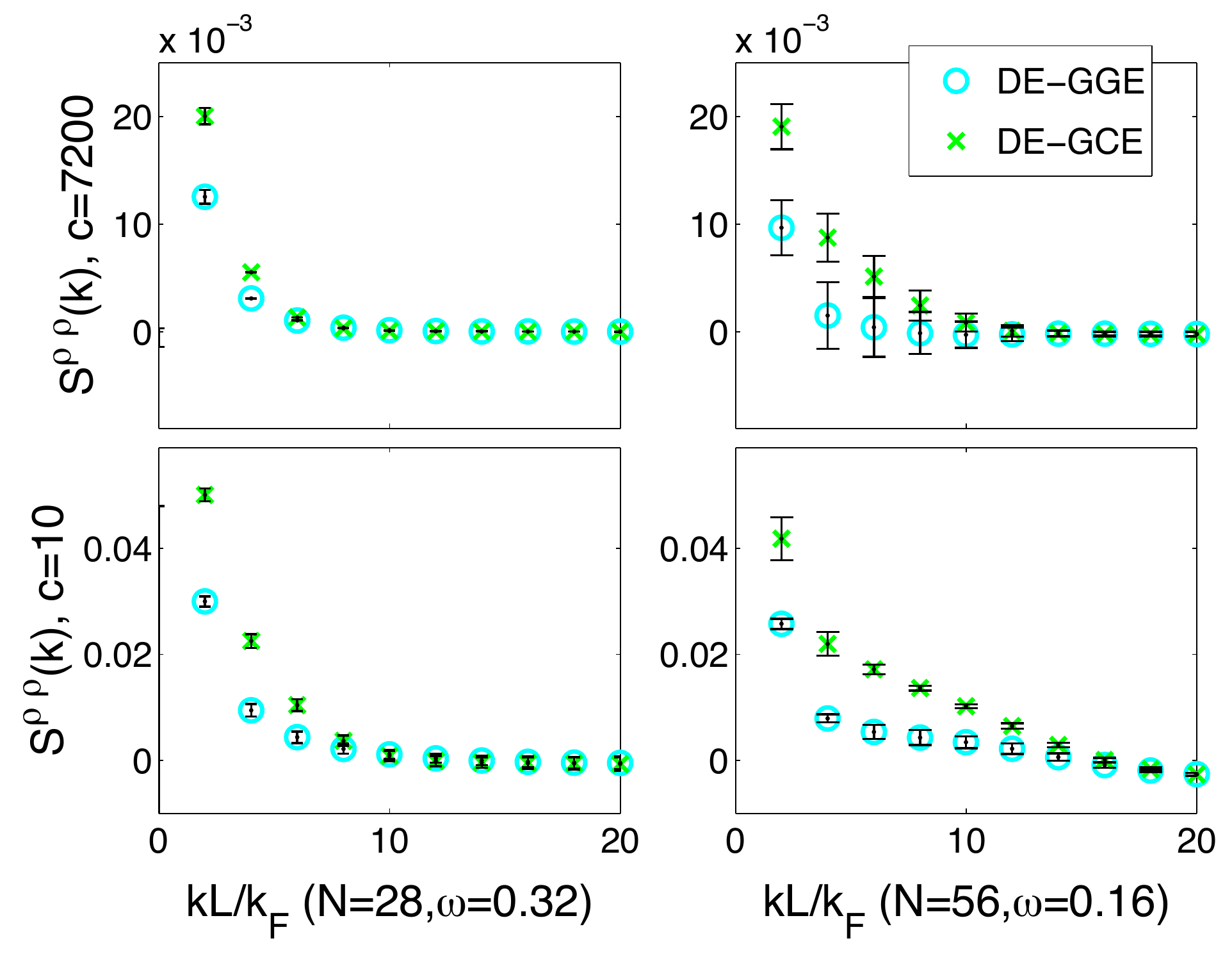}
\caption{The difference between the SSF as computed in the DE 
and as computed in the GGE and GCE for $N = L=28$ (left) and $N = L=56$ (right).
Top: The gas after release from a trap of strength $\omega = 0.32~(N=L=28)$ and $\omega = 0.16~ (N=L=56)$ for $c=7200$. 
Bottom: The gas after release from the same trap but for $c = 10$.
The GGE estimate is seen to be more accurate than the GCE throughout the range of values considered.}
\label{fig:SSFGE}
\end{figure}

\subsubsection{Contrasting the SSF in the DE, GGE and GCE}

In Figs. 8 and 9, we now contrast these results for the SSF in the DE 
with those obtained in the GGE and GCE.  We plot the results vs. momentum expressed in units of $kL/k_F$.
The particular form of the SSF in at least the diagonal ensemble at small $\omega$ then suggests that in doubling $N$ and $L$ while
halving $\omega$, the value of the SSF will double (see Eqn. \ref{srrde}).  

For both displayed values of the interaction strength, the agreement between the DE and GGE is very good, and much better than
between the DE and GCE.  This is true for the different trap strengths presented in both Figs. 8 and 9.  We note, however,
that the disagreement between the DE and GGE is larger for $c=10$ than
for $c=7200$ (Fig. 8).  

In terms of a finite size analysis, we note that for the data in Fig. 8, the DE-GGE curves maintain, roughly speaking, the same shape
between the $N=L=28$ and $N=L=56$.  Because of how the SSF is scaling with system size and our choice of units for momenta, this means
the difference between these two ensembles is decreasing as system size grows.  However the same cannot be said for the DE-GCE curves.
For the larger system size, the DE-GCE curves are notably more upturned at small momenta suggesting that in the infinite volume
limit, the two ensembles will yield different results for the SSF.  

This effect is, however, much less pronounced for the data in Fig. 9 where
a stronger trap is used.  Here the DE-GGE curve appears flatter for the larger system size data (N=L=56), while, the DE-GCE curve
appears much the same for the two different system sizes.  A more definable trend may be elusive here because of the larger uncertainties
associated with the larger trap values.

\subsection{Assessing the convergence of the computation of the MDF and SSF in the DE}

\subsubsection{MDF}
In computing the MDF in the DE for $N=L=56$ and $c=10$, we truncated the expression for the ground state to 7000 states
(for $N= L=14$, $100$ states were used; for $N = L=28$, $500$).  
The sum of the coefficients, $|c_s|^2$, over these 7000 states was $0.9884$ (for $N = L=14$ and $N = L=28$, respectively $0.9999$ and $0.9992$). 
The MDF itself can be checked using the sum rule $\frac{1}{L} \sum_k n_k = \frac{N}{L}$. 
The sum rule saturation for $N=L=56$ was $0.9690$, for $N=L=14$, $0.9995$, and for $N=L=28$, $0.9982$. 

For $N=L=56$ and $c=7200$, we kept $2000$ states, leading to the truncated sum $\sum_s|c_s|^2$ equalling
$0.9998$. The integrated MDF sum rule was saturated to $0.9900$.   For $N = L=14$, $100$ states were used, yielding saturations of $0.9999$ and $0.9992$
respectively.
For $N=L=28$, $500$ states were used, giving $0.9999$ and $0.9988$. 

\begin{figure}[t]
\includegraphics[width=0.4\textwidth,height=0.20\textwidth]{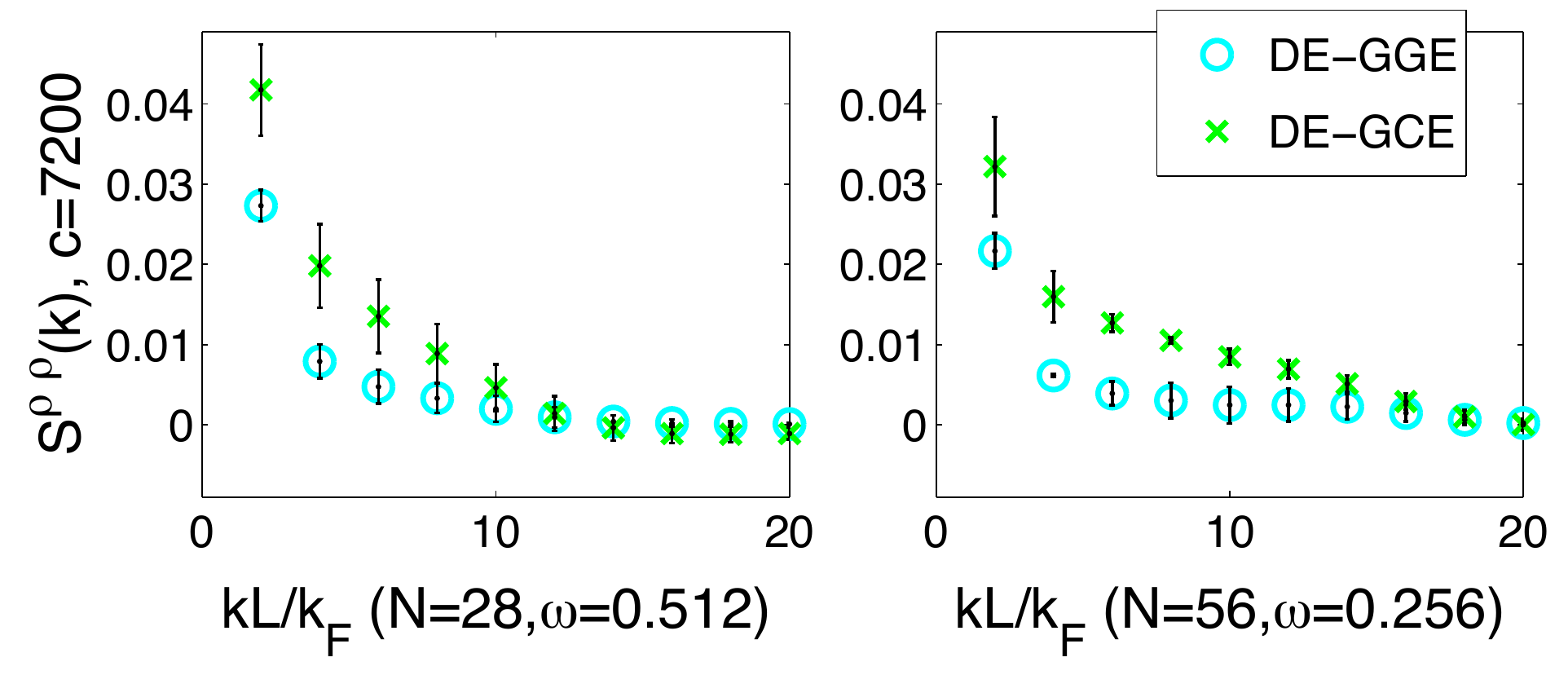}
\caption{The difference between the SSF as computed in the DE 
and as computed in the GGE and GCE for $N =L= 28$ (left) and $N =L= 56$ (right) for $c=7200$,
after release from traps of strengths $\omega = 0.512~(N=L=28)$ and $\omega = 0.256~(N=L=56)$. }
\label{fig:SSFGEw2}
\end{figure}

\subsubsection{SSF}
In computing the SSF in the DE for $c=10$ and $N=L=56$, we again truncated the expression for the ground
state to the same 7000 states.  
The f-sum rule $\int_{-\infty}^\infty \frac{d\omega}{2\pi} \omega S(k,\omega) = \frac{N}{L} k^2$
then provides a convenient check on the end result at any fixed momentum. 
This was saturated to $0.9884, 0.9766, 0.9451$ for $k = 2\pi/L, k_F$ and 
$2k_F$ by using ABACUS for excited states; the lack of saturation around $2k_F$ is visible in the top 
of Fig. \ref{fig:SSF}.

For the case of $c=7200$ we kept $1000$ states in the ground state 
expansion in the basis of Bethe states, yielding a truncated sum $\sum_s |c_s|^2$ equal to $0.9987$.  
This leads to a f-sum rule saturation 
for these three same momenta to the same value of $0.9987$.

For the other system sizes, we obtain the following saturations.  For $N=L=14$ and $c=10$,
we used 100 states saturating $0.9999$ of the wavefunction norm, and we obtained as the f-sum
saturations at the three wavevectors, $0.9999$, $0.9996$, and $0.9991$.
For $N=L=14$ and $c=7200$,
we used 100 states saturating $0.99999$ of the wavefunction norm, with
f-sum saturations, $0.99997$, $0.99997$, and $0.99997$.
For $N=L=28$ and $c=10$,
we used 500 states saturating $0.9992$ of the wavefunction norm with f-sum saturations, $0.9992$, $0.9964$, and $0.9915$.
And finally for $N=L=28$ and $c=7200$,
we used 500 states saturating $0.99997$ of the wavefunction norm, with
f-sum saturations, $0.99998$, $0.99998$, and $0.99998$.

\subsection{Calculation of Error Bars}

\subsubsection{Diagonal Ensemble}

The error bars for the MDF and the SSF in the diagonal ensemble (DE)
(as presented in Figs. 1 and 3 of the main text, and in Figs. 6-9 of the supplementary material)
were obtained according to the following scheme. The NRG gives a ground state in the form
$$
|GS\rangle = \sum_{s=1}^{M} c_s |s\rangle,
$$
where $|s\rangle$ are exact eigenstates of the untrapped gas.  This sum, formally, should be over {\it all} eigenstates
in the system (i.e. $M=\infty$).  In running the NRG to determine the coefficients $c_s$, we limit ourselves to a {\it finite} number of
states (approximately $M \approx 42 000$ for $N=L=56$).  The coefficients arrived at by the NRG always sum to 1, i.e. $\sum_{s=1}^{M}|c_s|^2=1$.
In computing the SSF or MDF in the DE, we further truncate this sum in order
to make the computation of the correlation function numerically manageable, for example keeping only the first $M'$ states ($M'=7000$
states for $N=L=56$).  The question is then what uncertainties these truncations introduce.  To get a feel for this for the
case $N=L=56$,
we compute the DE at two different truncation levels, the first using all $M'=7000$ states.  For this truncation the coefficients
$|c_s|^2$ sum to $\approx 0.99$ in all cases.  But we also look at a second, more severe, truncation where $\sum |c_s|^2 = 0.97$.
For both truncated wavefunctions, we renormalize the wavefunction so that it has unit norm.
We then ascribe the uncertainty to our computation of the DE to the difference between these two results.

We readily admit that this is a heuristic but we feel that it gives a ball park for the uncertainty and perhaps even
overestimates it.  

For the cases $N=L=14,28$, we feel the convergence of the sum $\sum |c_s|^2$ to $1$ is so rapid that the results for the MDF
and the SSF in the DE are completely converged and the error due to truncation is negligible (or at least smaller than the
symbol size used in the plots).
\vskip 40pt 

\subsubsection{Generalized Gibbs and Canonical Ensembles}

We first note that in computing the SSF (as well as the MDF) in the GGE and GCE, that while the
weights, $e^{-\varepsilon_{0-GGE}}$ and $e^{-\varepsilon_{0-GCE}}$, in these ensembles are computed without fixing
the particle number, when we compute the trace over states we only perform a partial trace involving those states
with the same particle number, $N$.  Thus the results presented in Figs. 6-9 of the supplementary material 
and 3 of the main text are computed, strictly
speaking, in fixed-density sub-ensembles. 

To obtain error bars for the GGE and GCE, we compute the SSF and MDF in two different sub-ensembles and
take the differences between these computations to obtain an (again heuristic) feel for the uncertainty.
For the $N=L=56$ case, the two (sub)-ensembles are the same collections of states used for determining uncertainties
in the DE, one sub-ensemble has a sum $\sum_s |c_s|^2 \approx 0.99$ while
the other satisfies $\sum_s |c_s|^2 \approx 0.97$.

For the $N=L=28$ case, the first sub-ensemble again satisfies $\sum_s |c_s|^2 \approx 0.99$ while
the second sub-ensemble is (differently from the $N=L=56$ case), is taken to be one half of the states of the first.
For this case, keeping instead only states that gave a saturation of $0.97$ led to too few states in the ensemble
to compute, even approximately, the SSF and MDF in the GGE and GCE.

\end{document}